\documentstyle[12pt,epsfig,color]{article}
\newcommand{\spacing}[1]{\renewcommand{\baselinestretch}{#1}\large\normalsize}
\spacing{1.5}
\begin{document}

\begin{center}
{\large\bf{HBT radii : Comparative studies on collision systems and beam energies}} \\
\bigskip

{\small
Debasish Das$^{a}$\footnote{email : debasish.das@saha.ac.in , dev.deba@gmail.com}
\medskip

$^a$Saha Institute of Nuclear Physics, HBNI, 1/AF, Bidhannagar, Kolkata 700064, India\\
}

\end{center}
\date{\today}
\begin{abstract}

Two-particle Hanbury-Brown-Twiss (HBT) interferometry is an important 
probe for understanding the space-time structure of particle emission 
sources in high energy heavy ion collisions. We present the comparative studies 
of HBT radii in Pb+Pb collisions at  $\sqrt{s_{\rm{NN}}}$ = 17.3 GeV with Au+Au collisions at 
$\sqrt{s_{\rm{NN}}}$ = 19.6 GeV. To further our understanding for this specific energy regime we 
also compare the HBT radii for Au+Au collisions at $\sqrt{s_{\rm{NN}}}$ = 19.6 GeV 
with Cu+Cu collisions at $\sqrt{s_{\rm{NN}}}$ = 22.4 GeV. We have found interesting 
similarity in the $R_{\rm out}/R_{\rm side}$ ratio with $m_{\rm T}$ across the collision 
systems while comparing the data for this specific energy zone which is interesting
as it acts as a bridge from SPS energy regime to the RHIC energy domain.

\end{abstract}


\section{\bf{Introduction}}

A phase transition from a hadronic state to a ``plasma'' of deconfined quarks 
and gluons when the energy density exceeds a critical value, 
is predicted from Quantum Chromo-Dynamics~(QCD).
The complicated structure of nuclear matter at low temperatures, 
where it is composed of a multitude of hadronic particles, baryons and mesons, 
is thus expected to give way at high temperatures to a plasma of weakly 
composed quarks and gluons, the ${\it Quark~-Gluon~Plasma}~(QGP)$. 
QGP is a thermalized system where the properties of the system are 
governed by the quark and gluon degrees of freedom~\cite{Karsch:2001cy}.

Understanding the deconfining phase transition in hadronic matter and
of the QGP properties is a challenging task. For systems created in the 
Relativistic Heavy Ion Collider~(RHIC) and Large Hadron collider~(LHC)
energy region with high temperatures and low baryo-chemical
potential, Lattice QCD calculations predict a cross-over transition
between the hadron gas and the QGP phase. Lattice QCD predicts a phase
transformation to a quark-gluon plasma at a temperature of approximately 
$T \approx 170~MeV (1~MeV \approx 1.1604 \times 10^{10}K)$ (~\cite{Karsch:2001cy})
corresponding to an energy density $\epsilon \approx 1~GeV/fm^{3}$, 
which is nearly an order of magnitude larger than normal nuclear matter.

Experimental studies in relativistic heavy ion physics aim to study the QCD
nature of matter under the conditions of extreme temperature and high energy density 
both at RHIC and at LHC. The discovery of the QGP can describe the system 
(governed by the quarks and gluons) in which the degrees of freedom are 
no more the colour neutral hadron states.

The equation of state (EoS) of nuclear matter enables us to understand 
the relationship between the pressure and the energy at a given net-baryon density. 
Phase transitions from the hadronic resonance gas phase to the color-deconfined QGP 
(see e.g.,~\cite{Spieles:1997ab,Bluhm:2007nu}), contribute to the 
changes of the EoS. The experimental measurements should also
be able to determine the physical characteristics of the transition, for example 
the critical temperature, the order of the phase transition, 
and the speed of the sound along with the nature of the quasi-particles. 
The EoS of hot and dense QCD matter is still not precisely understood.
Modern nuclear physics, has an important goal to explore the phase diagram of 
quark matter in various temperatures and baryon density so as to confirm the
existence of the new phase of quark matter~\cite{Fodor:2001pe,deForcrand:2006pv}.
 
The intermediate Super Proton Synchrotron (SPS) energy regime still 
remains interesting since the onset of deconfinement is expected to happen 
at those energies. Possibility of a critical endpoint~\cite{Lacey:2006bc,Liu:2015lse} and a
first-order phase transition is yet not excluded. Several
beam-energy dependent observables such as the particle ratios
\cite{Afanasiev:2002mx,Alt:2007fe}, the flow
\cite{Kolb:2000fha,Petersen:2006vm}, the HBT
parameters \cite{Rischke:1996em,Adamova:2002ff} show a
non-monotonic behaviour for which the interpretation still remains unclear. The Beam Energy Scan(BES) 
programs at RHIC, show that directed flow is strong for both the lowest and highest 
RHIC energies as shown by results from STAR experiment \cite{Adamczyk:2014ipa}. 
The net-proton $v_1(y)$ slope have a minimum between 11.5 and 19.6 GeV 
and changing sign twice between 7.7 and 39 GeV, which is quite contrary 
to the UrQMD transport model predictions for that energy regime.
The vanishing of directed flow when the expansion stops and its appearance 
when the matter has passed through the change is the ``latent heat'', 
where the predicted ``softest point disappearance'' of flow can become a possible  
signature of a first-order phase transition between hadronic matter and 
a deconfined QGP phase.

Assuming a first-order phase transition, there is a mixed phase 
of the QGP and hadronic gas. A slow-burning fireball is expected 
in the absence of pressure gradient, when the initial system is at 
rest in the mixed phase, and this leads to a time-delay in the system 
evolution~\cite{Rischke:1996em,Pra861,Ber89,Sof02}. Investigation of the time-delay signatures 
for the first-order phase transition is hence-forth a subject of interest.

Two-particle Hanbury-Brown-Twiss (HBT) interferometry is an important 
tool for detecting the space-time structure of particle emission 
sources in high energy heavy ion collisions~\cite{Won94,Wie99,Wei00}.  
The occurance of first-order phase transition between the QGP and hadronic matter, 
will lead to the time-delay of the system evolution and hence making the emission 
duration of particles more prolonged~\cite{Rischke:1996em,Pra861,Ber89,Sof02}.
As explained in these references~\cite{Rischke:1996em,Pra861,Ber89,Sof02} the three HBT radius parameters, 
$R_{\rm out}$, $R_{\rm side}$, $R_{\rm long}$, describe the dimensions of a Gaussian source 
in longitudinal co-moving system (LCMS) framework. The $R_{\rm out}/R_{\rm side}$ ratio can be 
related to the emission time~\cite{Rischke:1996em,Pra861,Ber89,Sof02}. 
We have explored in this paper the energy region of 17.3 GeV to 22.4 GeV 
through comparative studies of two-pion HBT radii. This energy region
has shown interesting results in STAR experiment \cite{Adamczyk:2014ipa}
for other correlation measurements (like flow).

\section{Results}

The intensity interferometry technique to measure sizes of stars~\cite{HanburyBrown:1956bqd} 
was formulated by Robert Hanbury Brown and Richard Twiss and also known as the ``Hanbury-Brown-Twiss (HBT) effect''.
Such technique was extended to particle physics~\cite{Goldhaber:1960sf} 
for understanding the angular distributions of pion pairs in $p\bar{p}$ annihilations, and thus 
the quantum statistics causing an enhancement in pairs with low relative momentum. 
In HBT analyses the method has henceforth evolved into a precision tool for measuring the 
space-time properties of the regions of homogeneity at kinetic freeze-out 
in heavy ion collisions~\cite{Lis05}.

\begin{figure}
\centering
\includegraphics[width=13.5cm]{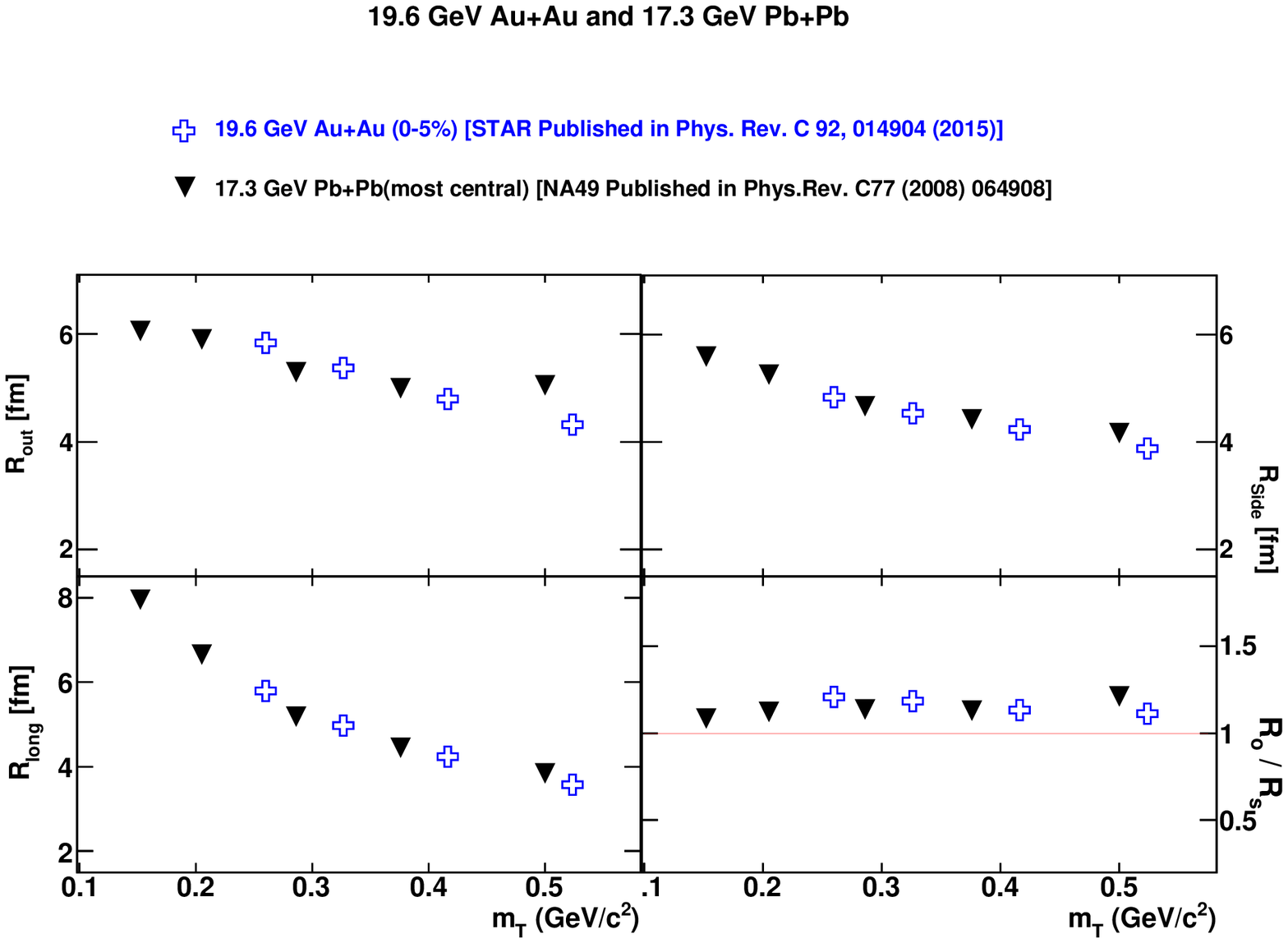}
\caption{The comparison of system size dependence in HBT radii of STAR Au+Au collisions at $\sqrt{s_{\rm{NN}}}$ = 19.6 GeV 
with NA49 Pb+Pb collisions for 17.3 GeV. Only statistical errors are shown for the top central data of both experiments.}
\end{figure}   

Two-pion interferometry yields HBT radii that describe the geometry of these
regions of homogeneity (regions that emit correlated pion pairs).
The HBT radii increases for more central collisions due 
to the increasing volume of the source and thus an example of how HBT can 
probe spatial sizes and shapes~\cite{Pratt:2008sz}.  
The decrease of HBT radii with mean pair transverse momentum, $k_{T}$(=$|\vec{p}_{\rm 1T}$ $+$ $\vec{p}_{\rm 2T}|$)$/$2),
has been due to transverse and longitudinal flow~\cite{Pratt:2008sz}. 
Flow causes space-momentum correlations since the size of the regions emitting the particles do not correspond 
to the entire fireball created in a relativistic heavy ion collision~\cite{Pratt:2008sz}.  

\begin{figure}
\centering
\includegraphics[width=13.5cm]{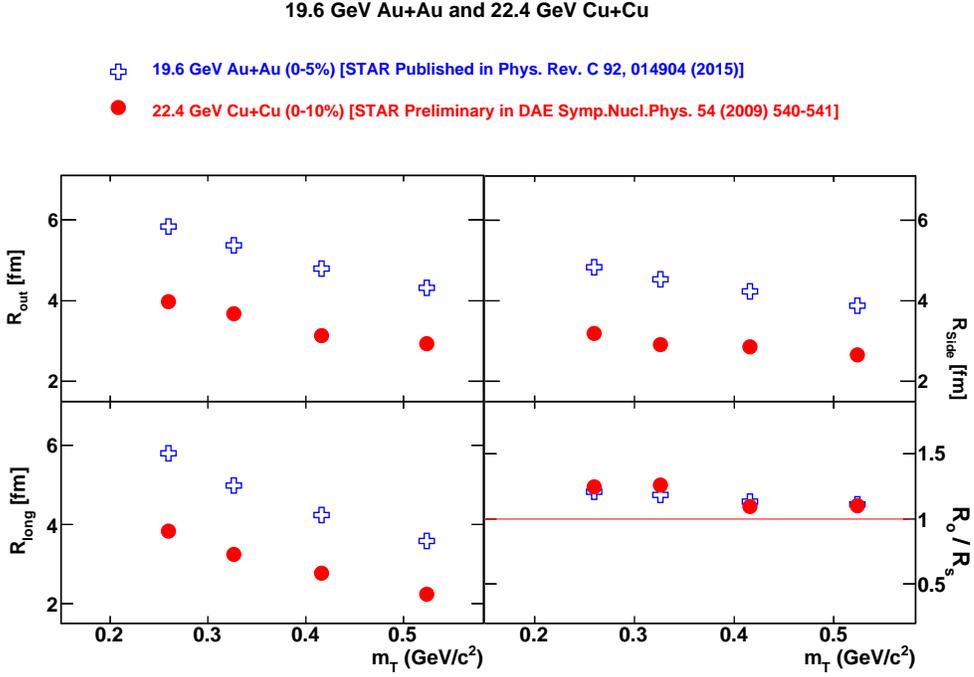}
\caption{The comparison of system size dependence in HBT radii of STAR Au+Au collisions at $\sqrt{s_{\rm{NN}}}$ = 19.6 GeV 
with Cu+Cu collisions at $\sqrt{s_{\rm{NN}}}$ = 22.4 GeV. Only statistical errors are shown for the top central data of both 
the Au+Au and Cu+Cu datasets.}
\end{figure}   

In this paper, the results of two-pion HBT analyses of Pb+Pb at 17.3 GeV 
from NA49 experiment~\cite{Alt:2007uj} are compared in Fig.1 and discussed 
with other STAR HBT results from Au+Au 19.6 GeV~\cite{Adamczyk:2014mxp}.
Fig.1 shows the HBT radii of SPS and RHIC collision species where for Pb+Pb 
17.3 GeV(NA49) and Au+Au 19.6 GeV(STAR) show similar trend for $R_{\rm side}$ 
and $R_{\rm long}$ with $m_{\rm T}$. For $R_{\rm out}$ the SPS data has a flatter
slope when compared with RHIC, but the $R_{\rm out}/R_{\rm side}$ ratios with 
$m_{\rm T}$ (= $\sqrt{k_{\rm T}^{2}+m_{\rm \pi}^{2}}$) are very similar for the top 
central data of both experiments. The $R_{\rm out}/R_{\rm side}$  ratios of NA49 and STAR show weak 
$m_{\rm T}$ dependence and have values close to unity.

The HBT radii from Au+Au 19.6 GeV and Cu+Cu 22.4 GeV both from 
STAR experiment are also included in this paper since they 
are different collision species with close by collision energies.
Reference~\cite{Das:2009wxi} explains the analysis methodology for Cu+Cu collisions at $\sqrt{s_{\rm{NN}}}$ = 22.4 GeV. 
In Fig.2 we present this comparison of two-pion HBT radii to include central (0-5$\%$) Au+Au collisions at 
$\sqrt{s_{\rm{NN}}}$ = 19.6 GeV and central (0-10$\%$) Cu+Cu collisions 
at $\sqrt{s_{\rm{NN}}}$ = 22.4 GeV from the STAR experiment.

The HBT radii for Cu+Cu collisions at $\sqrt{s_{\rm{NN}}}$ = 22.4 GeV are smaller 
than those for Au+Au collisions at $\sqrt{s_{\rm{NN}}}$ = 19.6 GeV. 
The variations of the $R_{\rm out}/R_{\rm side}$ ratios with $m_{\rm T}$ are similar for the
Au+Au and Cu+Cu collision data as we see in Fig.2. The ratios also show weak $m_{\rm T}$ dependence 
with the values close to unity.

\begin{figure}
\centering
\includegraphics[width=15cm]{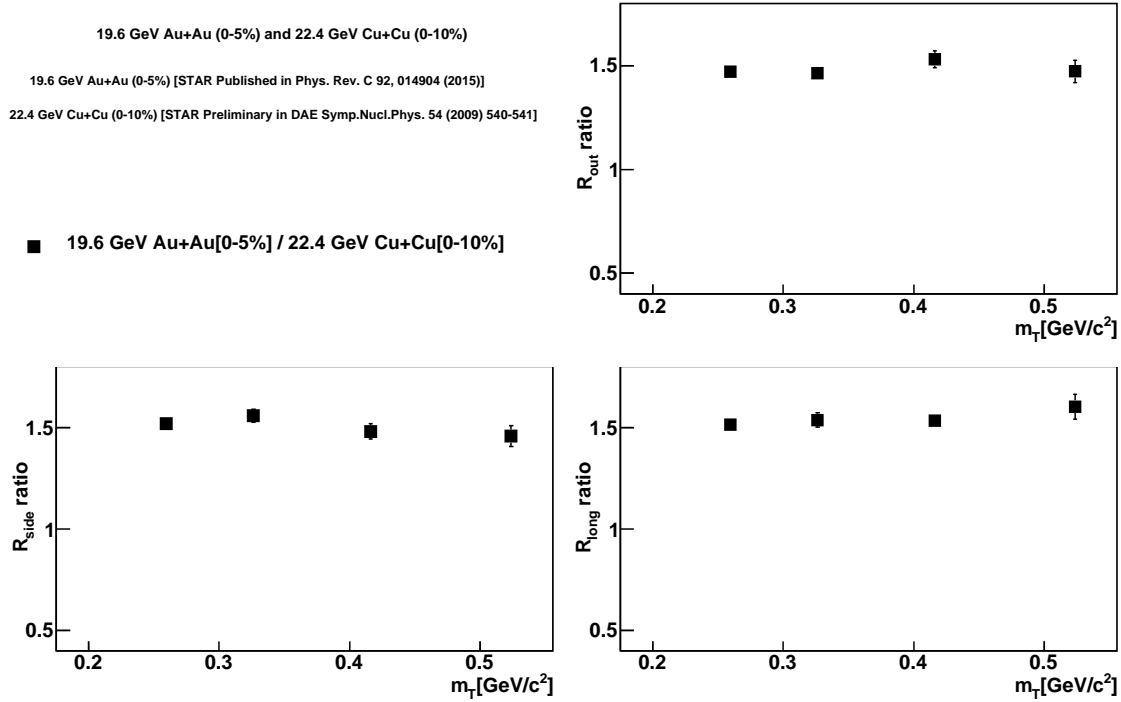}
\caption{Ratios of HBT radii at top centralities for Au+Au and Cu+Cu 
collisions at $\sqrt{s_{\rm{NN}}}$ = 19.6 and 22.4 GeV vs. $m_{T}$. Only statistical errors are shown for 
Au+Au collisions at $\sqrt{s_{\rm{NN}}}$ = 19.6 GeV and Cu+Cu collisions at $\sqrt{s_{\rm{NN}}}$ = 22.4 for 
their top central datasets.}
\end{figure}   

In Fig.3 we present the $m_{\rm T}$ dependences of the ratios of  
two-pion HBT radii for the most-central Au+Au at $\sqrt{s_{\rm{NN}}}$=19.6 GeV and Cu+Cu collisions at 
$\sqrt{s_{\rm{NN}}}$ = 22.4~GeV. Details about the Cu+Cu systems are explained in~\cite{Das:2009wxi} 
and references therein. As seen in Fig.3 the ratios of radii for Au+Au to Cu+Cu collisions are $\sim$1.5.  
Although we see that the individual HBT radii decrease significantly with increasing $m_{\rm T}$
but the the ratios in Fig.3 show that the HBT radii for 
Au+Au and Cu+Cu collisions at 19.6 GeV and 22.4 GeV share a common $m_{\rm T}$ dependence.
Such trends can be understood in terms of models~\cite{Cramer:2004ih,Miller:2005ji} 
where participant scaling is used to predict the HBT radii in Cu+Cu collisions
from the measured radii for Au+Au collisions at $\sqrt{s_{\rm{NN}}}$ = 200 GeV, 
assuming the radii are proportional to A$^{1/3}$, where A 
is the atomic mass number of the colliding nuclei.

\section{Summary}

The $R_{\rm out}/R_{\rm side}$ ratio is important since it is able to 
provide the information of the emission duration. We also know that the 
HBT radii are affected by transverse and longitudinal flow. The SPS energy regime is still
zone of interest where the recent flow results from STAR experiment \cite{Adamczyk:2014ipa}
(within 11.5 and 19.6 GeV) have shown some new and interesting features. When we 
compare the HBT (two-particle correlation) radii in Pb+Pb collisions at  $\sqrt{s_{\rm{NN}}}$ = 17.3 GeV with Au+Au collisions at 
$\sqrt{s_{\rm{NN}}}$ = 19.6 GeV we find very similar $R_{\rm out}/R_{\rm side}$ ratio with $m_{\rm T}$. 
To explore this interesting energy regime we have compared the HBT radii for Au+Au collisions at $\sqrt{s_{\rm{NN}}}$ = 19.6 GeV 
with Cu+Cu collisions at $\sqrt{s_{\rm{NN}}}$ = 22.4 GeV. The similarity in the $R_{\rm out}/R_{\rm side}$ 
ratio with $m_{\rm T}$ persists across the collision systems from SPS to RHIC energies and even 
in close by RHIC energies for Au+Au and Cu+Cu systems as well. The rise of the ratio $R_{\rm out}/R_{\rm side}$ 
with collision energy which was predicted~\cite{Rischke:1996em} due to a possible phase transition is not observed.
Such inferences establish that HBT radii $R_{\rm out}/R_{\rm side}$ ratios are very much
comparable and consistent across the different colliding species in (an exciting zone of interest of the RHIC BES program), 
the energy region of 17.3 GeV to 22.4 GeV.\\

{\bf Acknowledgements:}
Author D.D. acknowledges 
the facilities of Saha Institute of Nuclear Physics, Kolkata, India.\\
``The author declares that there is no conflict of interest regarding the publication of this paper.''
\\
\bibliography{apssamp}

\end{document}